\documentclass[fleqn,usenatbib,useAMS]{mnras}

\usepackage{newtxtext,newtxmath}

\usepackage[T1]{fontenc}

\DeclareRobustCommand{\VAN}[3]{#2}
\let\VANthebibliography\thebibliography
\def\thebibliography{\DeclareRobustCommand{\VAN}[3]{##3}\VANthebibliography}

\usepackage{graphicx}	
\usepackage{amsmath}	
\usepackage{orcidlink}

\title[White dwarf eccentricity]{White dwarf eccentricity fluctuation and dissipation by AGB convection}

\author[Y. Cohen et al.]{
Yair Cohen, Sivan Ginzburg$^{\orcidlink{0000-0002-3751-4553}}$,\thanks{E-mail: \href{mailto:sivan.ginzburg@mail.huji.ac.il}{sivan.ginzburg@mail.huji.ac.il}} Maya Levy, Tal Bar Shalom and Yoav Siman Tov
\\
Racah Institute of Physics, The Hebrew University, Jerusalem 91904, Israel
}

\date{Accepted XXX. Received YYY; in original form ZZZ}

\pubyear{2024}

\begin{document}
\label{firstpage}
\pagerange{\pageref{firstpage}--\pageref{lastpage}}
\maketitle

\begin{abstract}
Millisecond pulsars with white dwarf companions have typical eccentricities $e\sim 10^{-6}-10^{-3}$. The eccentricities of helium white dwarfs are explained well by applying the fluctuation--dissipation theorem to convective eddies in their red giant progenitors. We extend this theory to more massive carbon--oxygen (CO) white dwarfs with asymptotic giant branch (AGB) progenitors. Due to the radiation pressure in AGB stars, the dominant factor in determining the remnant white dwarf's eccentricity is the critical residual hydrogen envelope mass $m_{\rm env}$ required to inflate the star to giant proportions. Using a suite of \textsc{mesa} stellar evolution simulations with $\Delta m_{\rm c}=10^{-3}\,{\rm M}_{\sun}$ core-mass intervals, we resolved the AGB thermal pulses and found that the critical $m_{\rm env}\propto m_{\rm c}^{-6}$. The resulting eccentricity $e\sim 3\times 10^{-3}$ is almost independent of the remnant CO white dwarf's mass $m_{\rm c}$. Nearly all of the measured eccentricities lie below this robust theoretical limit, indicating that the eccentricity is damped during the common-envelope inspiral that follows the unstable Roche-lobe overflow of the AGB star. Specifically, we focused on white dwarfs with median masses $m_{\rm c}>0.6\,{\rm M}_{\sun}$. These massive white dwarfs begin their inspiral with practically identical orbital periods and eccentricities, eliminating any dependence on the initial conditions. For this sub-sample, we find an empirical relation $e\propto P^{3/2}$ between the final period and eccentricity that is much tighter than previous studies -- motivating theoretical work on the eccentricity evolution during the common envelope phase. The eccentricities of lower mass CO white dwarfs may be explained by alternative formation channels.
\end{abstract}

\begin{keywords}
stars: AGB and post-AGB -- binaries: general -- pulsars: general
 -- white dwarfs
\end{keywords}



\section{Introduction}\label{sec:intro}

Binary millisecond pulsars provide a unique opportunity to accurately measure orbits by timing the arrival of the orbiting pulsar's radio pulse \citep{LorimerKramer2004}. One example is the detection of Earth-mass and Moon-mass planets around pulsars \citep{WolszczanFrail1992,Wolszczan1994}. Another example is the measurement of eccentricities as low as $\sim 10^{-7}$ for white dwarf companions \citep{PhinneyKukarni1994,Freire2012}.

The orbits of low-mass white dwarfs ($\lesssim 0.45\,{\rm M}_{\sun}$) around pulsars are thought to be a fossil record of the white dwarf's formation process as a degenerate helium core inside a red giant progenitor star \citep{PhinneyKukarni1994}. Red giants burn hydrogen in a thin shell surrounding the degenerate core, resulting in a tight relation between the red giant's radius and the core's mass. As hydrogen burning keeps increasing the mass of the helium core, the red giant's hydrogen envelope expands to evacuate the increasing luminosity. At some point, the red giant overflows its Roche lobe around the pulsar and loses its envelope through stable mass transfer, leaving behind the helium white dwarf core with an orbital period that is a function of its mass \citep{RefsdalWeigert1969,RefsdalWeigert1970,RefsdalWeigert1971,Joss1987,Savonije1987,Rappaport1995,TaurisSavonije99}. 

The strong tides that the pulsar raises on the red giant's convective envelope damp the orbital eccentricity. However, the same convective eddies also perturb the orbit by stochastically altering the red giant's gravitational quadrupole moment. \citet{Phinney1992} showed that the residual eccentricity is determined by energy equipartition between the epicyclic motion and the convective eddies. Specifically, the low eccentricities observed for most helium white dwarf--millisecond pulsar binaries are set when the red giant's hydrogen envelope is reduced to $10^{-3}-10^{-2}\, {\rm M}_{\sun}$ (a few times the mass of the burning shell). At this point, the envelope drastically contracts by orders of magnitude, such that tides become too weak to change the eccentricity further. Similarly to the orbital period, the eccentricity also `freezes' at a value that is set by the degenerate core's mass, resulting in a distinct period--eccentricity relation. In recent years, several exceptions to the \citet{Phinney1992} relation were discovered; the origin of the much higher eccentricities in these systems is still an open question \citep{Antoniadis2014,FreireTauris2014,Jiang2015,HanLi2021,GinzburgChiang2022,Serylak2022,Tang2023,WangGong2023}.
      
More massive white dwarfs around millisecond pulsars are thought to have formed differently. When helium cores grow beyond $\approx 0.45\,{\rm M}_{\sun}$, the helium ignites and eventually a degenerate carbon--oxygen (CO) core builds up. When this CO core grows beyond $\approx 0.5\,{\rm M}_{\sun}$, it is surrounded by two thin shells -- hydrogen and helium -- that burn alternately in a series of `thermal pulses' \citep{Schwarzschild1965,Weigert1966,Kippenhahn2012}. This double shell structure is analogous to the single shell structure during the red giant branch (RGB), with the hydrogen envelope now expanding on the asymptotic giant branch (AGB). As the mass of the CO core grows, the giant's radius on the AGB exceeds the maximum RGB size, and the star may overflow its Roche lobe around the pulsar and initiate mass transfer. See \citet{Tauris2011conf} and \cite{Tauris2011,Tauris2012} for a detailed discussion of the binary stellar evolution leading to this and alternative formation scenarios. 

Due to mass loss by winds on the AGB, the progenitors of massive CO white dwarfs are believed to be initially more massive than their neutron star pulsar companions ($\gtrsim 2\,{\rm M}_{\sun}$), such that the mass transfer following Roche-lobe overflow tends to be unstable and lead to common envelope evolution \citep[e.g.][]{Paczynski1976,IbenLivio1993,Ivanova2013}. Unlike helium white dwarfs (which follow the \citealt{TaurisSavonije99} relation), the observed orbital periods of CO white dwarfs around millisecond pulsars are shorter than the Roche-lobe filling periods of their progenitor giants -- indicating an orbital inspiral within the common envelope \citep{PhinneyKukarni1994,vanDenHeuvel94,Hui2018}.

Despite their different formation histories, the observed eccentricities of helium and CO white dwarf--millisecond pulsar systems are similar, typically spanning $10^{-6}-10^{-3}$ \citep{Tauris2012,Hui2018,Parent2019}. While the eccentricities of helium white dwarfs are explained well by the \citet{Phinney1992} period--eccentricity relation, it is unclear what mechanism produces the very small -- but measurably non-zero -- eccentricities of CO white dwarfs.

Here, we extend the \citet{Phinney1992} theory to CO white dwarfs with AGB progenitor stars, and consider this mechanism as the possible origin of the eccentricities of such white dwarfs around millisecond pulsars, similarly to their helium white dwarf siblings. While the current periods of these binaries indicate that their orbits have shrunk since Roche-lobe overflow, we will test whether their eccentricities could have been set during this earlier stage. Specifically, we assume a two-stage process: the initial eccentricity is set by the \citet{Phinney1992} mechanism during Roche-lobe overflow, and then both the period and eccentricity evolve during a common envelope phase. At the very least, the \citet{Phinney1992} mechanism provides a clear quantitative limit -- without any adjustable parameters -- that can be compared to the measured eccentricities. This limit may serve as a theoretical anchor, given the uncertainty in the eccentricity evolution during the subsequent common envelope phase \citep{GlanzPerets2021,Szolgyen2022,Trani2022}.

The remainder of this paper is organized as follows. In Section \ref{sec:structure} we analyse the structure of RGB and AGB giant stars during mass transfer. In Section \ref{sec:orbital} we calculate the evolution of the orbital period and eccentricity during and after Roche-lobe overflow. We compare the theory to the observed periods and eccentricities of binary millisecond pulsars in Section \ref{sec:obs}. In Section \ref{sec:alternative} we briefly consider alternative formation paths.
We summarize and discuss our results in Section \ref{sec:summary}.

\section{Structure of giant stars}\label{sec:structure}

\subsection{Analytical scaling relations}\label{subsec:analytical}

We begin by deriving simplified analytical expressions for the stellar structure during mass transfer to the pulsar. As we demonstrate in Section \ref{subsec:numerical}, this description is a rather crude approximation, but it none the less provides valuable qualitative intuition.

\subsubsection{Burning shell}\label{sec:shell}

We assume a degenerate core (either helium or CO) with a mass $m_{\rm c}$ and a radius $r_{\rm c}\propto m_{\rm c}^{-1/3}$, surrounded by a non-degenerate hydrogen envelope with a mass $m_{\rm env}$ that extends to a radius $r_{\rm env}\gg r_{\rm c}$. At small radii $r$, where the core dominates the gravity, hydrostatic pressure equilibrium in the envelope dictates (see Section \ref{sec:envelope} for a more rigorous derivation)
\begin{equation}\label{eq:pressure}
    \frac{Gm_{\rm c}\rho}{r}\sim \frac{\rho}{\mu}kT+\frac{aT^4}{3},
\end{equation}
where $G$, $a$, and $k$ are the gravitational, radiation, and Boltzmann's constants respectively, $\rho$ is the gas density, $T$ is its temperature, and $\mu$ is the mean molecular weight. For helium cores, $m_{\rm c}<0.5\,{\rm M}_{\sun}$ and the radiation pressure is negligible, such that $T\propto m_{\rm c}/r$ \citep{Kippenhahn2012}. Assuming a volumetric energy production rate $\propto \rho^2T^\nu$, the fusion luminosity is given by
\begin{equation}\label{eq:l_nuc}
L\propto\int_{r_{\rm c}}{\rho^2 T^\nu}r^2{\rm d}r\propto\rho_{\rm c}^2T_{\rm c}^\nu r_{\rm c}^3,
\end{equation}
where $\rho_{\rm c}$ and $T_{\rm c}$ are evaluated on top of the core. The strong dependence on the temperature $\nu\gg 1$ ensures that the luminosity is dominated by a thin burning shell at $r\sim r_{\rm c}$. This luminosity is radiated away from the burning shell by diffusion
\begin{equation}\label{eq:l_diff}
L\sim\frac{\sigma T_{\rm c}^4 r_{\rm c}^2}{\tau}\propto\frac{T_{\rm c}^4 r_{\rm c}^2}{\kappa \rho_{\rm c}r_{\rm c}}\propto\frac{T_{\rm c}^4r_{\rm c}}{\rho_{\rm c}},    
\end{equation}
where $\sigma$ is the Stefan-Boltzmann constant and $\tau$ is the optical depth. The opacity $\kappa$ is given by a constant electron-scattering value at the relevant temperatures. By comparing equations \eqref{eq:l_nuc} and \eqref{eq:l_diff} we find
\begin{equation}\label{eq:l_low}
L\propto r_{\rm c}^{5/3}T_{\rm c}^{(\nu+8)/3}\propto m_{\rm c}^{(\nu+8)/3}r_{\rm c}^{-1-\nu/3}\propto m_{\rm c}^{3+4\nu/9}\approx m_{\rm c}^9,   
\end{equation}
\begin{equation}\label{eq:rho_mat}
\rho_{\rm c}\propto m_{\rm c}^{(4-\nu)/3}r_{\rm c}^{-2+\nu/3}\propto m_{\rm c}^{2-4\nu/9},    
\end{equation}
where $\nu\approx 13$ for the CNO cycle that dominates at the relevant temperatures. See \citet{RefsdalWeigert1970} and \citet{Kippenhahn2012} for a more rigorous derivation of equation \eqref{eq:l_low}. 

For more massive CO cores, with $m_{\rm c}\gtrsim 0.5\,{\rm M}_{\sun}$, radiation pressure becomes important and the mass--luminosity relation is shallower. By comparing the radiation dominated regime in equation \eqref{eq:pressure} with equation \eqref{eq:l_diff}, a much simpler relation is found \citep{Paczynski1970,Kippenhahn2012}:
\begin{equation}\label{eq:l_high}
L\propto\frac{T_{\rm c}^4r_{\rm c}}{\rho_{\rm c}}\propto m_{\rm c}.    
\end{equation}
Using equations \eqref{eq:l_nuc}, \eqref{eq:l_diff}, and \eqref{eq:l_high}, the density of the burning shell in the radiation dominated regime scales as
\begin{equation}\label{eq:rho_rad}
\rho_{\rm c}\propto m_{\rm c}^{(4-\nu)/(8+\nu)}r_{\rm c}^{(\nu-12)/(8+\nu)} \propto m_{\rm c}^{(24-4\nu)/(24+3\nu)}.
\end{equation}
We combine equations \eqref{eq:rho_mat} and \eqref{eq:rho_rad} and estimate the mass of the burning shell:

\begin{equation}\label{eq:m_sh}
m_{\rm sh}\sim \rho_{\rm c}r_{\rm c}^3\propto\begin{cases}
m_{\rm c}^{1-4\nu/9}\approx m_{\rm c}^{-5} & m_{\rm c}\ll 0.5\,{\rm M}_{\sun} \\
m_{\rm c}^{-7\nu/(24+3\nu)}\approx m_{\rm c}^{-13/9} & m_{\rm c}\gg 0.5\,{\rm M}_{\sun}
\end{cases}
\end{equation}
which is a strong function of the core mass $m_{\rm c}$ \citep[see also][]{Phinney1992}.

\subsubsection{Envelope}\label{sec:envelope}

In Section \ref{sec:shell} we assumed a radiative burning shell. In this section we extend our radiative model to the entire envelope in order to derive a simple analytical solution, even though red giant envelopes are actually mostly convective (in fact, we rely on this convection in Section \ref{sec:orbital}). The pressure is given by
\begin{equation}\label{eq:prs}
    p=\frac{\rho}{\mu}kT+\frac{aT^4}{3}=\frac{1}{\beta}\frac{\rho}{\mu}kT,
\end{equation}
where the ratio of matter to total pressure $\beta$ is assumed to be constant within the envelope. As long as the core dominates the mass of the star (we are interested in residual low-mass envelopes; see Section \ref{sec:intro}), hydrostatic equilibrium dictates
\begin{equation}\label{eq:hydrostat}
\frac{{\rm d}p}{{\rm d}r}=-\frac{Gm_{\rm c}\rho}{r^2}.    
\end{equation}
The radiative luminosity is given by the diffusion equation
\begin{equation}\label{eq:rad_trans}
L=-4\upi r^2 \frac{c}{3\kappa\rho}\frac{{\rm d}}{{\rm d}r}(aT^4)=-\frac{16\upi}{3}\frac{\sigma r^2}{\kappa\rho}\frac{{\rm d}T^4}{{\rm d}r},   
\end{equation}
where $c\equiv 4\sigma/a$ is the speed of light. By comparing equations \eqref{eq:hydrostat} and \eqref{eq:rad_trans}, demanding thermal equilibrium (i.e. $L$ is constant within the envelope), and assuming a constant $\kappa$ (valid for the inner layers of the envelope, where electron scattering dominates), we find
\begin{equation}\label{eq:power_law}
\frac{{\rm d}p}{{\rm d}r}\propto \frac{{\rm d}T^4}{{\rm d}r},    
\end{equation}
and therefore $p\propto T^4$ deep enough inside the envelope, where the external boundary conditions can be neglected. This justifies our constant $\beta$ assumption in equation \eqref{eq:prs}.
Using equations \eqref{eq:prs}, \eqref{eq:hydrostat}, and $p\propto T^4$, the temperature profile is given by
\begin{equation}\label{eq:tmp_prof}
\frac{{\rm d}T}{{\rm d}r}=\frac{1}{4}\frac{T}{P}\frac{{\rm d}p}{{\rm d}r}=-\frac{\beta}{4}\frac{Gm_{\rm c}\mu}{kr^2},    
\end{equation}
such that deep enough inside the envelope $\rho\propto T^3\propto r^{-3}$. Equation \eqref{eq:pressure} can be reproduced by combining equations \eqref{eq:prs} and \eqref{eq:tmp_prof}.
We can now integrate the density profile to calculate the envelope's mass
\begin{equation}\label{eq:menv_ln}
\begin{split}
m_{\rm env}&=\int_{r_{\rm c}}^{r_{\rm env}}{4\upi r^2\rho(r){\rm d}r}=\int_{r_{\rm c}}^{r_{\rm env}}{4\upi r^2\rho_{\rm c}\left(\frac{r_{\rm c}}{r}\right)^3{\rm d}r}
\\
&\sim m_{\rm sh}\ln\frac{r_{\rm env}}{r_{\rm c}},        
\end{split}
\end{equation}
or alternatively
\begin{equation}
r_{\rm env}\sim r_{\rm c}\exp\left(\frac{m_{\rm env}}{m_{\rm sh}}\right),   
\end{equation}
which is valid for $m_{\rm sh}<m_{\rm env}<m_{\rm env}^{\rm crit}$, with $m_{\rm env}^{\rm crit}$ defined below. 

Equations \eqref{eq:l_low} and \eqref{eq:l_high} indicate that the luminosity of the giant in this regime (when $r_{\rm env}\gg r_{\rm c}$) depends solely on $m_{\rm c}$, and is independent of $m_{\rm env}$. Therefore, if we hypothetically engulf the core with progressively more massive envelopes (starting from $m_{\rm env}=m_{\rm sh}$), we find that the effective black-body temperature decreases with increasing $m_{\rm env}$, scaling as\footnote{This explains the almost horizontal track of post-AGB stars in the Hertzsprung--Russell diagram \citep{Blocker2001}.}
\begin{equation}
T_{\rm eff}=\left(\frac{L}{4\upi r_{\rm env}^2\sigma}\right)^{1/4}\propto L^{1/4}(m_{\rm c})r_{\rm env}^{-1/2}.  
\end{equation}
However, when the effective temperature drops below $\approx 5\times 10^3\,{\rm K}$, the photosphere's ${\rm H}^{-}$ opacity becomes extremely sensitive to $T_{\rm eff}$, such that $T_{\rm eff}$ becomes approximately constant, following the Hayashi line \citep{RefsdalWeigert1970}. For such massive envelopes, $r_{\rm env}$ becomes independent of $m_{\rm env}$ and reaches its maximum value
\begin{equation}\label{eq:renv}
r_{\rm env}^{\rm max}\propto L^{1/2}\propto\begin{cases}
m_{\rm c}^{9/2} & m_{\rm c}\ll 0.5\,{\rm M}_{\sun} \\
m_{\rm c}^{1/2} & m_{\rm c}\gg 0.5\,{\rm M}_{\sun}
\end{cases}
\end{equation}
with a more accurate fit given by \citet{Rappaport1995}. 

In other words, giants maintain their large radii (which are independent of $m_{\rm env}$), as long as $m_{\rm env}>m_{\rm env}^{\rm crit}$, which is several times the mass of the burning shell. More specifically, using equation \eqref{eq:menv_ln}
\begin{equation}\label{eq:m_crit}
m_{\rm env}^{\rm crit}(m_{\rm c})\sim m_{\rm sh}(m_{\rm c})\ln\frac{r_{\rm env}^{\rm max}(m_{\rm c})}{r_{\rm c}(m_{\rm c})}.
\end{equation}
Once the envelope drops below this mass, it contracts exponentially, eventually reaching a radius comparable to the underlying white dwarf core. 

\subsection{Numerical calculations}\label{subsec:numerical}

We evolved $1\,{\rm M}_{\sun}$ and $2\,{\rm M}_{\sun}$ stars on the RGB and AGB respectively using the stellar evolution code \textsc{mesa}, version r23.05.1 \citep{Paxton2011,Paxton2013,Paxton2015,Paxton2018,Paxton2019,Jermyn2023}. Our input files closely follow \citet{Cinquegrana2022} and \citet{Navarro2023}, who focused on the thermally pulsating AGB phase. As explained in Section \ref{subsec:analytical}, the structure of a giant star depends almost exclusively on the mass of its underlying degenerate core. The choice of the total stellar mass (i.e. core plus envelope) is therefore arbitrary and made for numerical convenience. For this reason, we artificially disable stellar winds such that our $1\,{\rm M}_{\sun}$ and $2\,{\rm M}_{\sun}$ progenitors scan a wide range of core masses $m_{\rm c}$ as they evolve.

\begin{figure}
\includegraphics[width=\columnwidth]{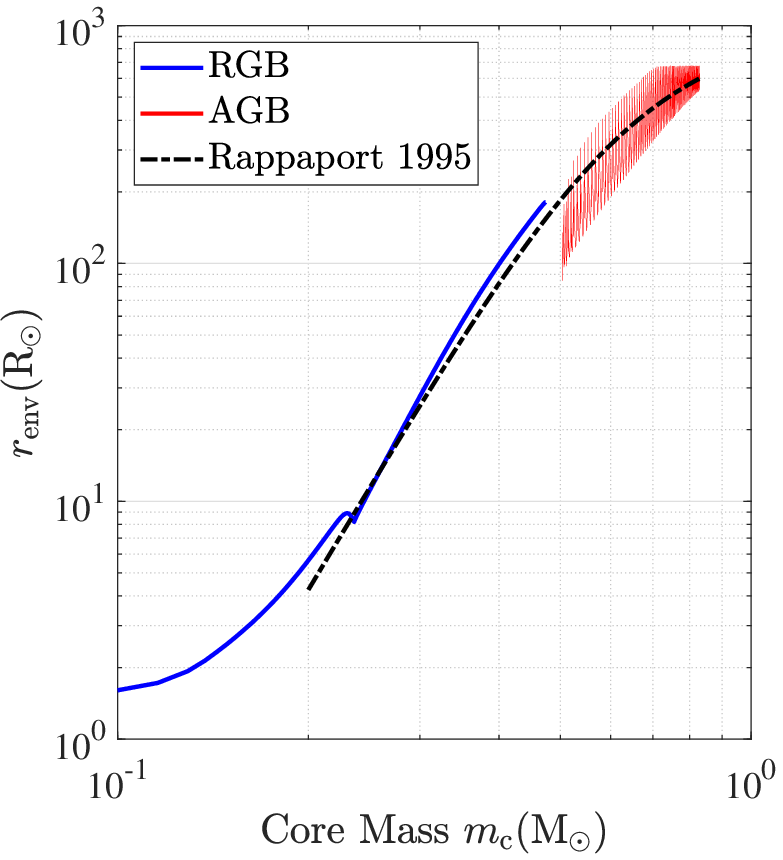}
\caption{The radius $r_{\rm env}$ of a giant star's envelope on the RGB and AGB as a function of its core mass $m_{\rm c}$, obtained by evolving $1\,{\rm M}_{\sun}$ (for the RGB) and $2\,{\rm M}_{\sun}$ (for the AGB) stars in \textsc{mesa} and disabling stellar winds. On the RGB the core is composed of helium, whereas on the AGB it is composed of carbon and oxygen. The radius on the AGB oscillates due to thermal pulses. We fit $r_{\rm env}(m_{\rm c})$ on both branches with the approximate equation \eqref{eq:rap95}, following \citet{Rappaport1995}.}
\label{fig:radius}
\end{figure}

\begin{figure}
\includegraphics[width=\columnwidth]{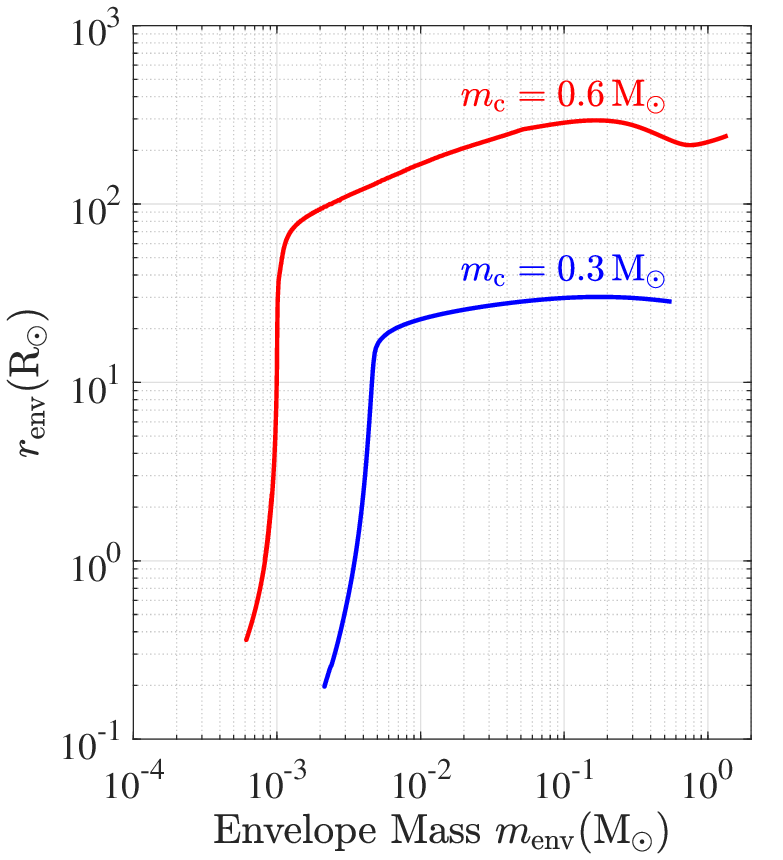}
\caption{The radii of giant stars $r_{\rm env}$ as a function of their hydrogen envelope mass $m_{\rm env}$ for different core masses $m_{\rm c}=0.3\,{\rm M}_{\sun}$ (on the RGB) and $m_{\rm c}=0.6\,{\rm M}_{\sun}$ (on the AGB). As long as $m_{\rm env}$ is larger than a critical value $m_{\rm env}^{\rm crit}(m_{\rm c})$, the giant's radius $r_{\rm env}$ is roughly constant (and given by Fig. \ref{fig:radius}). Once the envelope's mass drops below this critical value, its radius shrinks by orders of magnitude, disabling tidal interactions with the pulsar companion. Specifically, convective eddies stop pumping the orbit's eccentricity when $m_{\rm env}\approx m_{\rm env}^{\rm crit}$ \citep{Phinney1992}.}
\label{fig:menv}
\end{figure}

\begin{figure}
\includegraphics[width=\columnwidth]{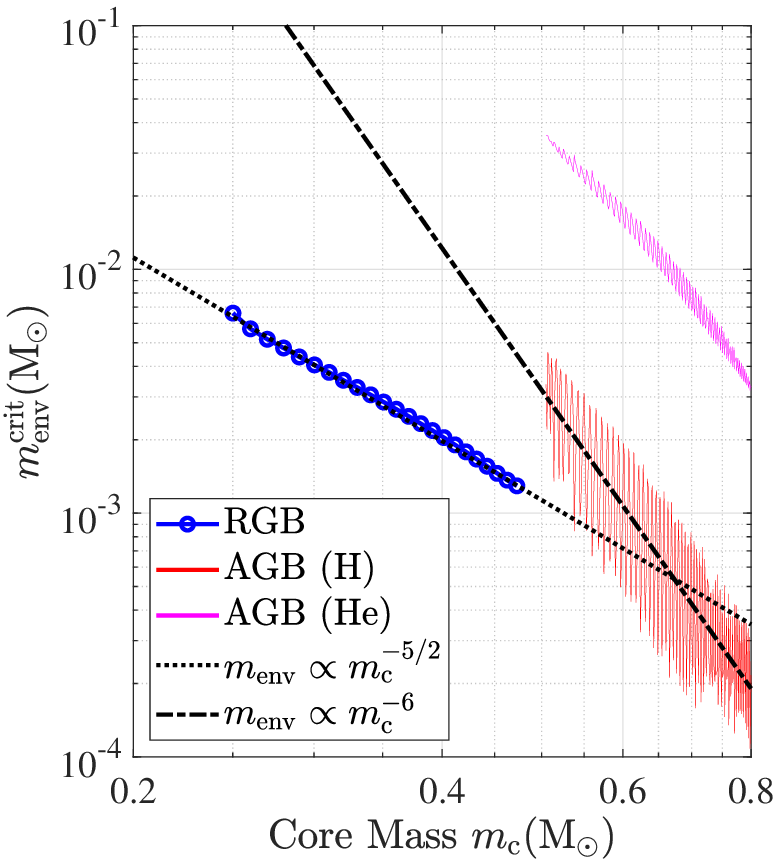}
\caption{The critical hydrogen envelope mass $m_{\rm env}^{\rm crit}$ at which $r_{\rm env}$ drops by a factor of 10 below the giant's original radius (we are not sensitive to this arbitrary definition thanks to the sharp drop, as seen in Fig. \ref{fig:menv}). We computed $m_{\rm env}$ as a function of the helium core mass $m_{\rm c}$ on the RGB (blue circles) and as a function of the CO core mass $m_{\rm c}$ on the AGB (red line, which oscillates due to the thermal pulses). Approximate power-law fits are provided for both branches. For completeness, we also plot the mass of the helium shell on the AGB (magenta line). Despite being more massive than the hydrogen envelope, this shell is much thinner such that it does not tidally interact with the pulsar.}
\label{fig:mcrit}
\end{figure}

Fig. \ref{fig:radius} shows the expansion of the star's envelope as the core mass grows, in accordance with equation \eqref{eq:renv}. We follow \citet{Rappaport1995}, with slight modifications, and fit $r_{\rm env}(m_{\rm c})$ with
\begin{equation}\label{eq:rap95}
r_{\rm env}\approx 6\times 10^3 \frac{(m_{\rm c}/{\rm M}_{\sun})^{9/2}}{1+7(m_{\rm c}/{\rm M}_{\sun})^4}\,{\rm R}_{\sun},    
\end{equation}
which reproduces both limits of equation \eqref{eq:renv}. In Fig. \ref{fig:menv} we gradually remove the hydrogen envelope (at an almost fixed $m_{\rm c}$) and demonstrate how giants maintain their radii until $m_{\rm env}$ drops below $m_{\rm env}^{\rm crit}(m_{\rm c})$. In Fig. \ref{fig:mcrit} we repeat this exercise for $\Delta m_{\rm c}=10^{-2}\,{\rm M}_{\sun}$ intervals on the RGB and $\Delta m_{\rm c}=10^{-3}\,{\rm M}_{\sun}$ intervals on the AGB to compute $m_{\rm env}^{\rm crit}(m_{\rm c})$. While on the RGB $m_{\rm env}$ is well defined, on the AGB the degenerate core is surrounded by two nested shells -- helium and hydrogen. The radius of the star is dominated by the outer hydrogen envelope, so we define $m_{\rm env}$ as the mass of the hydrogen layer alone (i.e. without the inner thin helium shell); this extended mass is the one that is relevant for the tidal interaction with the companion pulsar (Section \ref{sec:orbital}).

Our numerically computed $m_{\rm env}^{\rm crit}$ decreases steeply with increasing $m_{\rm c}$ (Fig. \ref{fig:mcrit}).\footnote{Extrapolating our power-law fit reveals that $m_{\rm env}^{\rm crit}=m_{\rm c}$ for $m_0\approx 0.09\,{\rm M}_{\sun}$, which is close to the hydrogen burning limit separating main-sequence stars from degenerate objects. Plugging $m_0$ in equation \eqref{eq:rap95} also roughly reproduces the main-sequence/degenerate radius at this mass $r_0\approx 0.1\,{\rm R}_{\sun}$. This is not a coincidence: for $m_0$, the temperature above the degenerate core $T_{\rm c}$ is equal to the hydrogen burning temperature in main-sequence stars, such that the radius $r_{\rm c}$, density $\rho_{\rm c}$, and luminosity $L$ of the burning shell are similar to those of a main-sequence star at this mass. In other words, $r_{\rm env}\sim r_{\rm c}\sim r_0$ and $m_{\rm sh}\sim m_{\rm c}\sim m_0$ -- providing a useful normalization point for red giants.} While this qualitative behaviour is predicted by the analytical equations \eqref{eq:m_sh} and \eqref{eq:m_crit}, our fitted power-laws differ considerably from equation \eqref{eq:m_sh}. Possible sources for this discrepancy include the variation of the nuclear burning exponent $\nu$ with temperature (and therefore with $m_{\rm c}$), the logarithmic factor in equation \eqref{eq:m_crit}, which also depends on $m_{\rm c}$, and the marginal degeneracy of the hot core, which leads to deviations from $r_{\rm c}\propto m_{\rm c}^{-1/3}$ \citep{RefsdalWeigert1970}. In addition, our single shell analytical theory (Section \ref{subsec:analytical}) does not capture the alternating burning of the double shell structure on the AGB and the resulting thermal pulses \citep[e.g.][]{Belloni2024}.

\section{Orbital evolution}\label{sec:orbital}

In this section we briefly describe the \citet{Phinney1992} model, and extend it to AGB stars using the results of Section \ref{sec:structure}. According to \citet{Phinney1992}, the orbital period $P$ and eccentricity $e$ are determined close to the moment of Roche-lobe detachment, when $m_{\rm env}\approx m_{\rm env}^{\rm crit}\ll m_{\rm c}$. For simplicity, we approximate $m_{\rm c}\ll M$, where $M$ is the mass of the pulsar, and assume that all white dwarfs orbit pulsars with roughly the same mass $M$. These simplifications will enable us to derive useful analytical power-law relations.

The orbital period at which the giant star fills its Roche lobe is given by its mean density
\begin{equation}\label{eq:roche}
P^2\sim\frac{R^3}{GM}\sim\frac{r_{\rm env}^3}{Gm_{\rm c}},    
\end{equation}
where $R$ is the distance between the giant and the pulsar, and $r_{\rm env}\approx r_{\rm env}^{\rm max}$ as long as $m_{\rm env}>m_{\rm env}^{\rm crit}$. When $m_{\rm env}$ drops below this critical value, the giant quickly contracts and detaches from its Roche lobe. The orbital period at the moment of detachment is given by 
\begin{equation}\label{eq:period_cases}
P\propto r_{\rm env}^{3/2}m_{\rm c}^{-1/2}\propto
\begin{cases}
m_{\rm c}^{25/4} & m_{\rm c}\ll 0.5\,{\rm M}_{\sun} \\
m_{\rm c}^{1/4} & m_{\rm c}\gg 0.5\,{\rm M}_{\sun}
\end{cases}
\end{equation}
where we have substituted $r_{\rm env}$ from equation \eqref{eq:renv}; see \citet{Rappaport1995} for a similar expression and \citet{TaurisSavonije99} for a more accurate computation in the $m< 0.45\,{\rm M}_{\sun}$ regime.

The envelopes of giant stars are mostly convective \citep[e.g.][]{Kippenhahn2012}. To evacuate the nuclear burning luminosity $L$, the convective flux satisfies
\begin{equation}\label{eq:conv_flux}
F=\frac{1}{2}\rho_{\rm env}v_{\rm conv}^3=\sigma T_{\rm eff}^4,    
\end{equation}
where $\rho_{\rm env}\sim m_{\rm env}r_{\rm env}^{-3}$ is the envelope's mean density and $v_{\rm conv}$ is the typical convective velocity. The convection slightly changes the giant's gravitational quadrupole moment (the monopole and dipole are fixed by the conservation of mass and momentum), perturbing the binary orbit. \citet{Phinney1992} showed that the epicyclic motion (associated with the eccentricity around the equilibrium circular orbit) reaches energy equipartition with the convective eddies \citep[see also][]{GinzburgChiang2022}:
\begin{equation}\label{eq:equi}
\frac{GMm_{\rm c}}{2R}e^2=\frac{1}{2}m_{\rm env}v_{\rm conv}^2.    
\end{equation}
Using equations \eqref{eq:conv_flux} and \eqref{eq:equi}, the eccentricity scales as
\begin{equation}
e^2\propto Rm_{\rm c}^{-1}m_{\rm env}^{1/3}r_{\rm env}^2\propto P^{2/3}m_{\rm c}^{-1}m_{\rm env}^{1/3}r_{\rm env}^2,    
\end{equation}
where we assumed an almost constant $T_{\rm eff}$ on the Hayashi line and used Kepler's third law. Specifically, if the giant fills its Roche lobe, then its orbital period is given by equation \eqref{eq:roche}, such that
\begin{equation}\label{eq:ecc}
e\propto P\left(\frac{m_{\rm env}}{m_{\rm c}}\right)^{1/6}\propto m_{\rm c}^{-2/3}m_{\rm env}^{1/6}r_{\rm env}^{3/2}.    
\end{equation}

The tidal circularization time-scale on which this eccentricity equilibrium is established scales as $(r_{\rm env}/R)^{-8}$ \citep{Zahn1977,Phinney1992,GinzburgChiang2022}.\footnote{The same convective eddies that perturb the orbit and excite the eccentricity also damp the eccentricity by dissipating the energy stored in the tidal bulges that the pulsar raises. This is a manifestation of the fluctuation--dissipation theorem \citep{Phinney1992}.} 
As long as the giant star fills its Roche lobe such that $r_{\rm env}/R\sim (m_{\rm c}/M)^{1/3}$, this time-scale is short enough to enforce equation \eqref{eq:ecc} as the star evolves (we check this in Appendix \ref{sec:timescales}). However, once the envelope's mass is reduced to $m_{\rm env}\approx m_{\rm env}^{\rm crit}$ and $r_{\rm env}$ contracts by a factor of a few, the circularization time-scale exceeds the contraction time-scale. At this point, the contraction `runs away' and tides are no longer able to adjust the eccentricity. As a result, the eccentricity freezes when $m_{\rm env}\approx m_{\rm env}^{\rm crit}$ and $r_{\rm env}\sim r_{\rm env}^{\rm max}$. By plugging $r_{\rm env}$ from equation \eqref{eq:renv} in equation \eqref{eq:ecc} we find that the eccentricity scales approximately as
\begin{equation}
e\propto
\begin{cases}
m_{\rm c}^6m_{\rm env}^{1/6} & m_{\rm c}\ll 0.5\,{\rm M}_{\sun} \\
m_{\rm c}^{1/12}m_{\rm env}^{1/6} & m_{\rm c}\gg 0.5\,{\rm M}_{\sun}
\end{cases}
\end{equation}
We now use our power-law fits for $m_{\rm env}^{\rm crit}(m_{\rm c})$ from Fig. \ref{fig:mcrit}, and find that for $m_{\rm c}\ll 0.5\,{\rm M}_{\sun}$ the dependence on $m_{\rm env}$ can be neglected, whereas for $m_{\rm c}\gg 0.5\,{\rm M}_{\sun}$ this term actually dominates the scaling:
\begin{equation}\label{eq:ecc_cases_simple}
e\propto
\begin{cases}
m_{\rm c}^6 & m_{\rm c}\ll 0.5\,{\rm M}_{\sun} \\
m_{\rm c}^{-1} & m_{\rm c}\gg 0.5\,{\rm M}_{\sun}
\end{cases}
\end{equation}
where we approximate both powers for simplicity.

Combining the low-mass limit (RGB) of equations \eqref{eq:period_cases} and \eqref{eq:ecc_cases_simple} approximately yields the famous relation $e\propto P$ for helium white dwarfs \citep{Phinney1992}. As we show in Section \ref{sec:obs}, the orbits of more massive CO white dwarfs have shrunk after Roche-lobe overflow, such that their observed periods are below the prediction of equation \eqref{eq:period_cases}. Without modelling this inspiral (which is beyond the scope of the paper), our theory cannot predict an analogous $e(P)$ relation for $m_{\rm c}>0.5\,{\rm M}_{\sun}$. Never the less, we can still use the $e(m_{\rm c})$ relation implied by equation \eqref{eq:ecc_cases_simple} to estimate the role of convective eddies in setting the observed eccentricities of CO white dwarfs. 

\section{Observations}\label{sec:obs}

\begin{figure}
\includegraphics[width=\columnwidth]{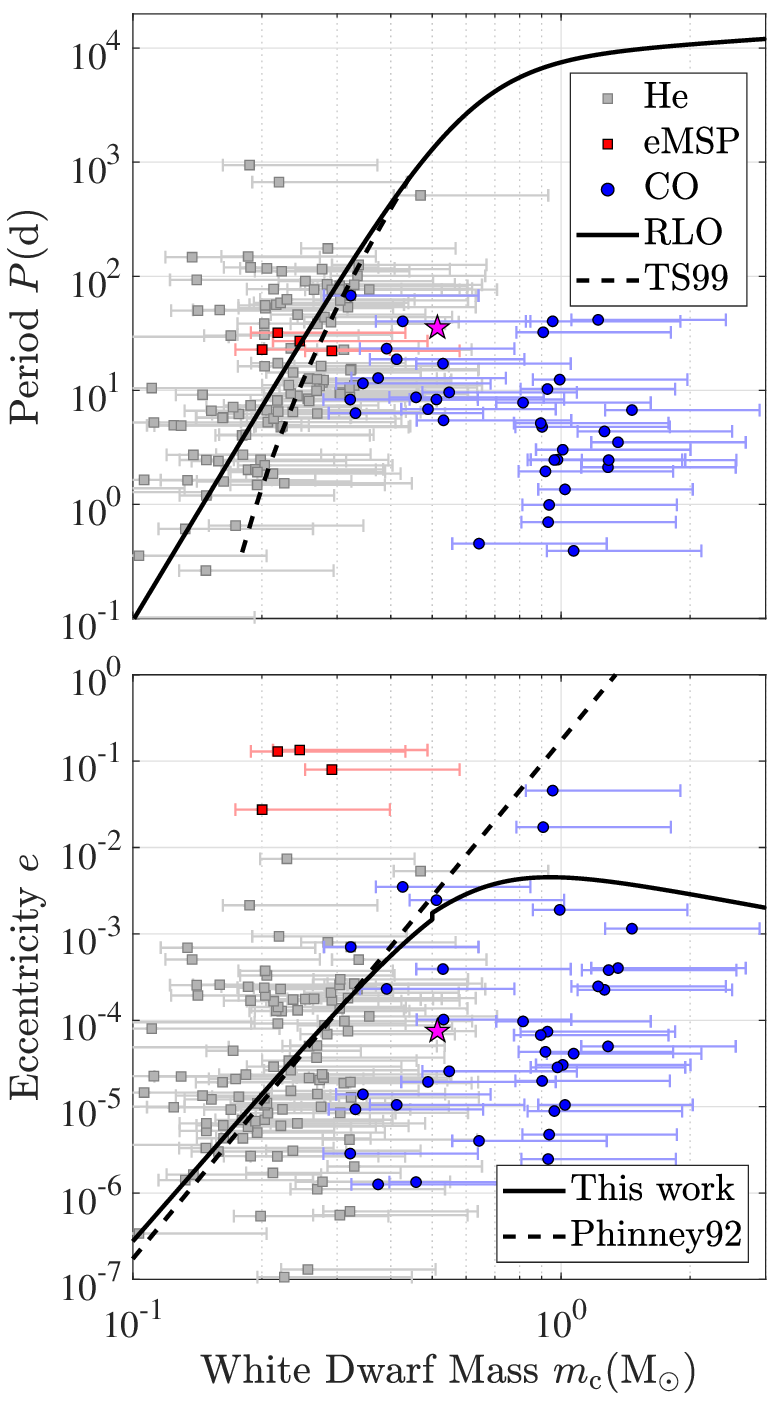}
\caption{Observed field binary millisecond pulsars from the ATNF Pulsar Catalogue with helium (squares) and CO/ONe (blue circles) white dwarf companions. \textit{Top panel:} unlike helium white dwarfs, the orbital periods of CO white dwarfs are inconsistent with the Roche-lobe overflow of their progenitors (solid line), with a radius $r_{\rm env}$ given by equation \eqref{eq:rap95}. The dashed line marks the detailed evolutionary calculations of \citet{TaurisSavonije99}. \textit{Bottom panel:} the eccentricities of most helium white dwarfs are set by convection in their progenitors' envelopes when they detach from their Roche lobe (\citealt{Phinney1992}; the exceptional eMSPs are coloured in red). We extend this theory to CO white dwarfs using equation \eqref{eq:ecc}, with envelope masses $m_{\rm env}$ fitted from Fig. \ref{fig:mcrit}. This limit (solid line) provides the initial conditions for the subsequent inspiral from Roche-lobe overflow to the current periods. The magenta star markers indicate the results of an alternative formation path -- Case A/B Roche-lobe overflow -- which we discuss in Section \ref{sec:alternative}.} 
\label{fig:obs}
\end{figure}

In Fig. \ref{fig:obs} we plot the observed millisecond pulsars (spin periods shorter than 50 ms) with white dwarf companions from the ATNF Pulsar Catalogue \url{http://www.atnf.csiro.au/research/pulsar/psrcat} \citep{Manchester2005}, version 2.1.1 (April 2024). We exclude pulsars that are associated with globular clusters because their eccentricities could have been excited by dynamical interactions \citep{RasioHeggie95}. The markers indicate the median white dwarf mass (which we identify with the progenitor's core mass $m_{\rm c}$), and the error bars span from the minimum mass to the 90th percentile (inclination angle of $25.8^{\circ}$).

\begin{figure}
\includegraphics[width=\columnwidth]{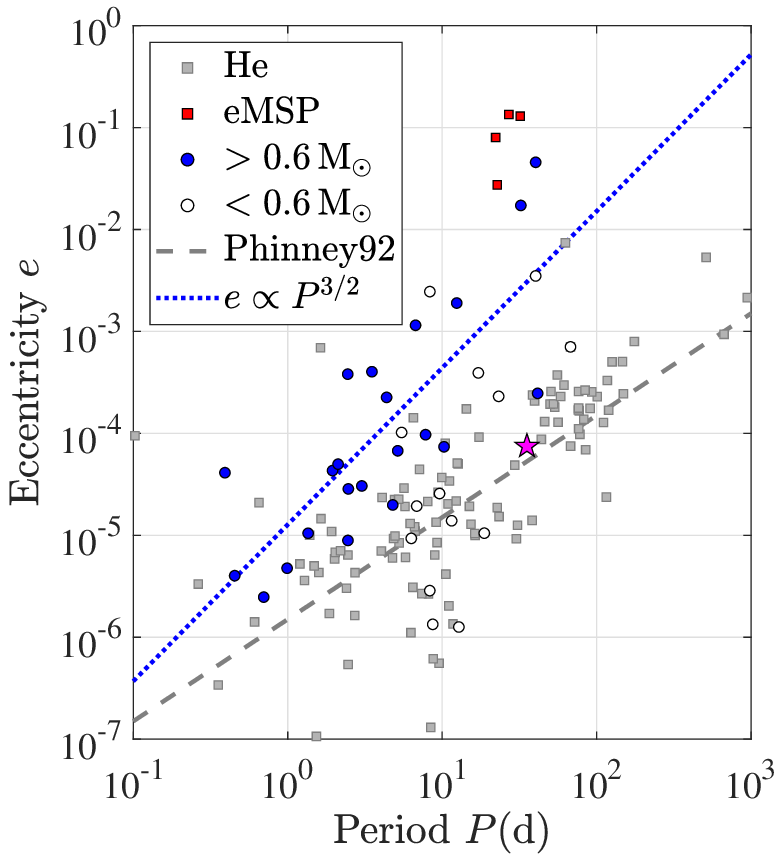}
\caption{Similar to Fig. \ref{fig:obs}, but in the period--eccentricity plane. We also separate the CO white dwarfs (circles) according to their median mass. White dwarfs with $m_{\rm c}>0.6\,{\rm M}_{\sun}$ begin their common-envelope inspiral with similar orbital periods and eccentricities (solid lines in Fig. \ref{fig:obs}), eliminating a potential dependence on these initial conditions. Our best power-law fit for their final $e(P)$ relation (dotted line) may therefore teach us about the eccentricity evolution during common envelope. 
The eccentricities of helium white dwarfs (except for eMSPs) follow the \citet{Phinney1992} relation  $e\propto P$ (dashed line).}
\label{fig:ep}
\end{figure}

We divide the observed population into helium white dwarfs and CO white dwarfs (the latter group possibly containing also oxygen--neon ONe white dwarfs). We overlay the observations with the orbital period during Roche-lobe overflow, which is calculated \citep[similarly to][]{Rappaport1995} from the envelope's radius, given by equation \eqref{eq:rap95}. The measured periods of helium white dwarfs are similar to the periods of their Roche-lobe filling progenitors, consistent with stable mass transfer \citep[see also][]{PhinneyKukarni1994,Rappaport1995}. The measured eccentricities of most of these helium white dwarfs are explained well by convective perturbations in the progenitor's envelope during Roche-lobe detachment \citep{Phinney1992}. A notable exception is the recently discovered class of eccentric millisecond pulsars (eMSPs), which are clustered at orbital periods of $\approx 20-30$ d \citep[see][and references therein]{GinzburgChiang2022}. The \citet{Phinney1992} relation is represented in Fig. \ref{fig:obs} by $e\propto m_{\rm c}^6$ (the dashed line in the bottom panel), but due to the uncertainty in $m_{\rm c}$, it is usually depicted in the $e$--$P$ plane instead \citep[Fig. \ref{fig:ep}, see also][]{PhinneyKukarni1994,GinzburgChiang2022}. 

The observed orbital periods of CO white dwarfs, on the other hand, are orders of magnitude shorter than the periods at which their progenitors filled their Roche lobe -- indicating an inspiral following an unstable mass transfer episode (i.e. a common envelope event).
We estimate the eccentricity at the onset of the inspiral by extending the \citet{Phinney1992} model using equation \eqref{eq:ecc}: by substituting $r_{\rm env}$ from equation \eqref{eq:rap95} and $m_{\rm env}$ from the analytical fits in Fig. \ref{fig:mcrit}, we derive a more accurate version of equation \eqref{eq:ecc_cases_simple}, which is given by the solid line in the bottom panel of Fig. \ref{fig:obs}. We normalize our $e(m_{\rm c})$ relation, as well as the $e\propto m_{\rm c}^6$ dashed line, to coincide with \citet{Phinney1992} at an orbital period of 100 d ($m_{\rm c}\approx 0.3\,{\rm M}_{\sun}$). We emphasize that this normalization is not a freely adjustable fitting parameter, but it was rather calculated theoretically by \cite{Phinney1992} using a more accurate version of our Section \ref{sec:orbital}. As a consequence, our theory has no free parameters, and it implies that convection excites eccentricities $e\sim 3\times 10^{-3}$ for all CO white dwarfs during Roche-lobe overflow, regardless of their mass (because of the nearly flat $e$--$m_{\rm c}$ relation at these masses). 

Almost all of the measured eccentricities of CO white dwarfs lie below our theoretical limit (bottom panel of Fig. \ref{fig:obs}), indicating that the orbital inspiral is accompanied by circularization. Although a theoretical description of such circularization is beyond the scope of this paper, we might gain some empirical insight by studying the observed $e(P)$ relation, which we plot in Fig. \ref{fig:ep}. In particular, CO white dwarfs with $m_{\rm c}\gtrsim 0.6\,{\rm M}_{\sun}$ offer a unique opportunity: as seen in Fig. \ref{fig:obs}, such massive white dwarfs begin their inspiral with practically identical periods and eccentricities -- eliminating any potential dependence on the initial conditions. Our best power-law fit for CO white dwarfs with median masses $m_{\rm c}>0.6\,{\rm M}_{\sun}$ is approximately $e\propto P^{3/2}$ (Fig. \ref{fig:ep}), which is much shallower and tighter than \citet{Hui2018}, who unlike us did not decompose their sample into massive and low-mass CO white dwarfs. 

\section{Alternative formation paths}\label{sec:alternative}

So far we have focused on `Case C' mass transfer from AGB stars harbouring degenerate CO cores, leading to common envelope inspiral. However, there are alternative paths to form CO white dwarf--millisecond pulsar binaries that avoid both the AGB and the common envelope phases \citep[see][for a concise review]{Tauris2011conf}. For a certain region of the parameter space (i.e. initial donor mass, orbital period, and neutron star mass), mass transfer that is initiated on the main sequence (Case A) or the RGB (Case B) remains stable and the donor ignites its helium core after detaching from the Roche lobe, leaving behind a CO white dwarf remnant \citep{Tauris2000,ShaoLi2012,Misra2020}.

\begin{figure}
\includegraphics[width=\columnwidth]{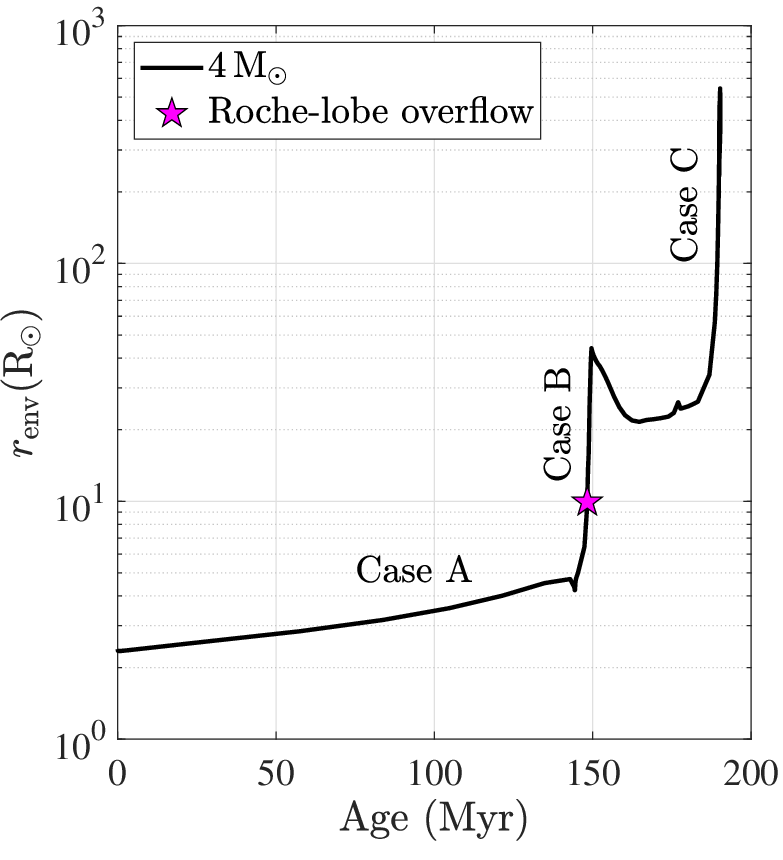}
\caption{Single stellar evolution of a $4\,{\rm M}_{\sun}$ star. When in a binary, such a star may overflow its Roche lobe while on the main sequence (Case A), RGB (Case B), or AGB (Case C) -- depending on the initial orbital period $P$.
As a specific example, the magenta star marker indicates the overflow moment for $P=5\,{\rm d}$ around an $M=1.8\,{\rm M}_{\sun}$ neutron star. The binary evolution of the overflowing star in this example is plotted in Fig. \ref{fig:binary}.}
\label{fig:single}
\end{figure}

\begin{figure}
\includegraphics[width=\columnwidth]{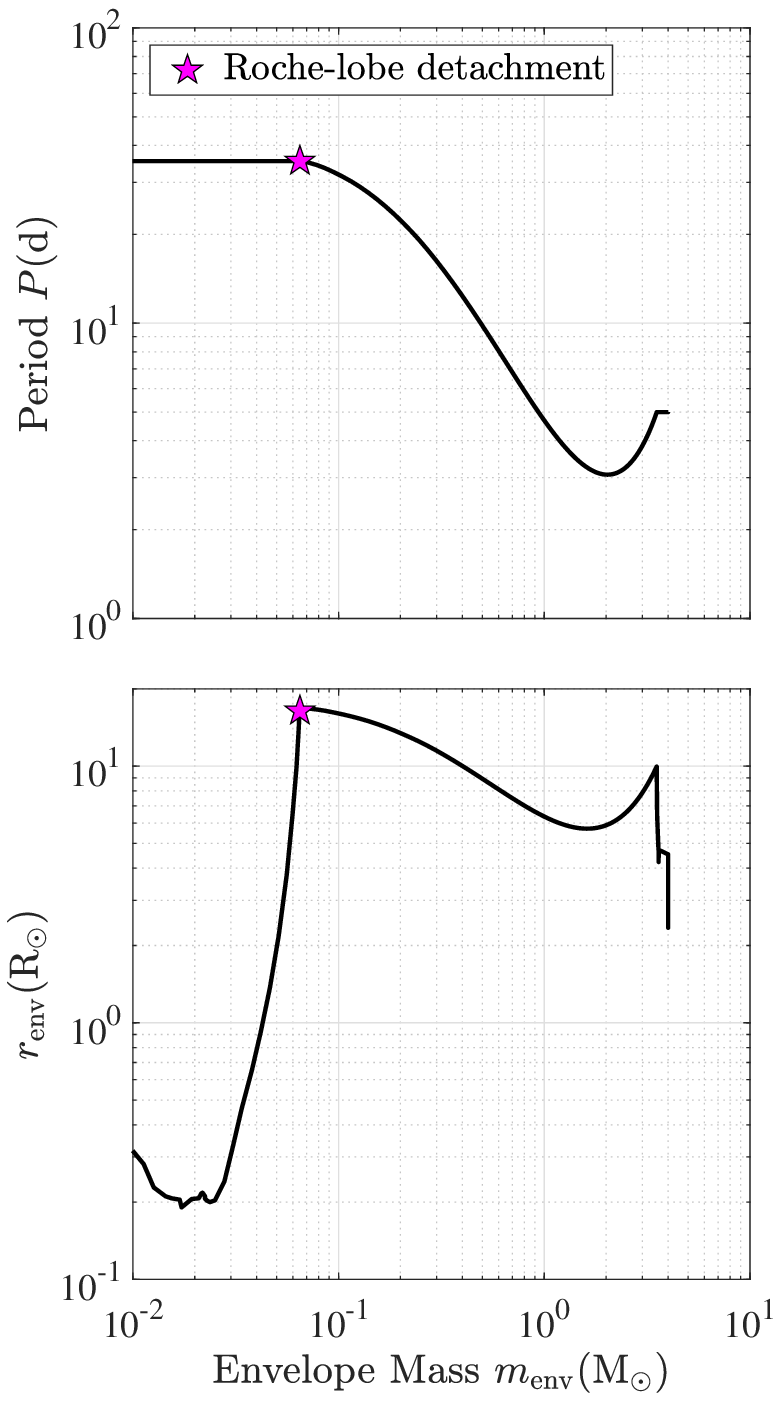}
\caption{Binary stellar evolution of an initially $4\,{\rm M}_{\sun}$ star on a $P=5\,{\rm d}$ orbit around an $M=1.8\,{\rm M}_{\sun}$ neutron star. After Case B Roche-lobe overflow (see Fig. \ref{fig:single}), the star detaches (magenta star markers) and contracts, eventually forming an $m_{\rm c}\approx 0.5\,{\rm M}_{\sun}$ CO white dwarf. This detachment sets the white dwarf's final orbital period $P$ (top panel) and eccentricity $e$ according to equation \eqref{eq:ecc}, with the critical $m_{\rm env}$ inferred from the bottom panel. These results are compared to the observations in Fig. \ref{fig:obs}.}
\label{fig:binary}
\end{figure}

In Figs \ref{fig:single} and \ref{fig:binary} we examine a typical example of such a Case A/B formation scenario: an initially $4\,{\rm M}_{\sun}$ donor with an initial orbital period $P=5\,{\rm d}$ around an $M=1.8\,{\rm M}_{\sun}$ neutron star. We evolved this system using the \texttt{star\_plus\_point\_mass} binary test suite in \textsc{mesa}.
Following Case B Roche-lobe overflow, the donor detaches from its Roche lobe at an orbital period $P\approx 35\,{\rm d}$, when its envelope is reduced to a critical mass $m_{\rm env}\approx 6\times 10^{-2}\,{\rm M}_{\sun}$. Similarly to \citet{Phinney1992} and our Section \ref{sec:orbital}, the subsequent contraction of the donor star well below its Roche lobe shuts off tidal interactions and freezes the eccentricity evolution. We apply equation \eqref{eq:ecc} with our regular normalization (which, again, is not a free parameter) and find $e\approx 7\times 10^{-5}$ for the $m_{\rm c}\approx 0.5\,{\rm M}_{\sun}$ CO white dwarf remnant. As expected and verified in Fig. \ref{fig:ep}, this value is very close to the original \citet{Phinney1992} $e\propto P$ relation, which was derived under similar assumptions; the small difference stems from the relatively weak dependence on $m_{\rm env}/m_{\rm c}$, which was neglected by \citet{Phinney1992}.  

As seen in Fig. \ref{fig:obs}, this alternative formation path produces CO white dwarfs with periods and eccentricities that are close to some of the observations. Unlike Case C overflow, our computed $P$ and $e$ (magenta star markers) actually represent the final values of these observables, because the system avoids a common envelope inspiral. Case A/B overflow therefore provides a potential alternative explanation for the measured CO white dwarf eccentricities. This scenario is mainly relevant for lower mass CO white dwarfs \citep{Tauris2000,ShaoLi2012,Misra2020}, i.e. the empty circles in Fig. \ref{fig:ep}. The higher mass white dwarfs on which we focus in this paper (the filled circles in Fig. \ref{fig:ep}) probably formed through Case C mass transfer followed by common envelope inspiral, as assumed in the previous sections \citep[see fig. 4 of][]{Tauris2011conf}. Our promising results in this section for a typical example of Case A/B evolution motivate a more extensive investigation of the mass, period, and eccentricity ranges of white dwarfs that are formed in this scenario. We defer such a survey of the parameter space to future work, and focus here on the AGB path.

\section{Summary and discussion}\label{sec:summary}

One of the impressive successes of stellar evolution theory is explaining the orbits of millisecond pulsars with helium white dwarf companions. Stable Roche-lobe overflow of the white dwarf's red giant (RGB) progenitor sets a clear relation $P\propto m_{\rm c}^6$ between the orbital period and the white dwarf's mass, which is consistent with the observations \citep{Joss1987,PhinneyKukarni1994,Rappaport1995,TaurisSavonije99}. The measured orbital eccentricities follow a similar correlation $e\propto P\propto m_{\rm c}^6$, which is set by convective motion of mass in the red giant's envelope \citep{Phinney1992}. Specifically, the orbital evolution ($P$ and $e$) freezes when the red giant's hydrogen envelope is reduced to a critical mass $m_{\rm env}^{\rm crit}$ (a few times the mass of the burning shell), at which its radius $r_{\rm env}$ contracts by orders of magnitude inside the Roche lobe. The success of this theory has been recently challenged by the discovery of several eccentric millisecond pulsars (eMSPs) that follow the $P(m_{\rm c})$ relation, but are orders of magnitude more eccentric than predicted by \citet{Phinney1992}. \citet{GinzburgChiang2022} argued that these eMSPs could be explained by a resonance between the orbital period and the convective turnover time, but several other mechanisms have been proposed as well \citep{Antoniadis2014,FreireTauris2014,Jiang2015,HanLi2021,Tang2023,WangGong2023}.   

In this paper, we turned our attention to a similar class of binary millisecond pulsars with heavier CO white dwarf companions. The formation history of these systems is less obvious than their helium counterparts \citep{Tauris2011conf,Tauris2011,Tauris2012}. 
Specifically, their current orbital periods are orders of magnitude shorter than the Roche-lobe filling periods of their presumed AGB progenitor stars, possibly indicating an inspiral within a common envelope following unstable Roche-lobe overflow. Surprisingly, despite this different evolution, the typical measured eccentricities of these two classes are rather similar ($e\sim 10^{-6}-10^{-3}$), leading us to speculate whether they were excited by the same mechanism. Other models might face a fine-tuning challenge in trying to explain such tiny -- but finite -- eccentricities.

To answer this question, we extended the \citet{Phinney1992} model to AGB stars that initially fill their Roche lobe. Due to the increasing role of radiation pressure in their interiors, the $P(m_{\rm c})$ relation of these stars flattens \citep{Rappaport1995}, such that the mass of the envelope $m_{\rm env}^{\rm crit}$ -- rather than that of the core $m_{\rm c}$ -- becomes the dominant factor in determining the eccentricity. By running a suite of \textsc{mesa} simulations with $\Delta m_{\rm c}=10^{-3}\,{\rm M}_{\sun}$ intervals, we resolved the thermal pulses on the AGB and found that $m_{\rm env}^{\rm crit}\propto m_{\rm c}^{-6}$. For the most massive white dwarfs, $m_{\rm env}^{\rm crit}$ is so low that the lengthening circularization time-scale $t_{\rm circ}$ freezes the eccentricity even before the progenitor star contracts within its Roche lobe (Appendix \ref{sec:timescales}). Quantitatively, AGB convection yields an approximately constant $e\sim 3\times 10^{-3}$, which is almost independent of $m_{\rm c}$.

We compared our theoretical prediction -- which has no free parameters -- to the observations, finding that almost all CO white dwarf--millisecond pulsar binaries have lower eccentricities than predicted. This indicates that orbital eccentricities are damped, rather than excited, during the subsequent common envelope inspiral phase. Quantitatively, we found an empirical relation $e\propto P^{3/2}$ between the final eccentricities and periods of white dwarfs with median masses $m_{\rm c}>0.6\,{\rm M}_{\sun}$. Presumably thanks to the almost uniform Roche-lobe overflow $e$ and $P$ \textit{initial} conditions in this mass range, our \textit{final} $e$--$P$ relation is much tighter than \citet{Hui2018}, who considered less massive CO white dwarfs as well. This observed correlation motivates future theoretical work on the eccentricity evolution during the common envelope phase \citep[see also][]{Dermine2013,GlanzPerets2021,Szolgyen2022,Trani2022}. 

For the less massive CO white dwarfs -- with median masses $m_{\rm c}<0.6\,{\rm M}_{\sun}$ -- we also considered an alternative scenario of Case A/B stable overflow before their giant progenitors have the chance to ascend the AGB \citep{Tauris2000,ShaoLi2012,Misra2020}. We computed a typical binary stellar evolution track of an $m_{\rm c}\approx 0.5\,{\rm M}_{\sun}$ CO white dwarf formed through this channel, reproducing $P$ and $e$ of some of the observations without an uncertain common envelope phase. Our preliminary result motivates a more comprehensive examination of the eccentricity fluctuation and dissipation in this alternative formation path.

Although we focused here on radio observations of white dwarfs around millisecond pulsars, our results may be applied to overflowing AGB stars (and their white dwarf remnants) orbiting other types of stars as well. Specifically, the \textit{Gaia} mission has recently uncovered a population of white dwarf--main sequence wide binaries that could have formed similarly \citep{Hallakoun2024,Shahaf2024,Yamaguchi2024}. \citet{Yamaguchi2024_sample} studied a sample of such stars with white dwarf masses $m_{\rm c}>0.6\,{\rm M}_{\sun}$, similar to our sample of massive white dwarfs in Fig. \ref{fig:ep}. Interestingly, the measured eccentricities of these stars $e\sim 10^{-2}-10^{-1}$ lie above our upper limit for stable mass transfer (Fig. \ref{fig:obs}), but they fit our empirical $e\propto P^{3/2}$ relation for the post common envelope eccentricity (Fig. \ref{fig:ep}).
Another recent example is the class of self-lensing binaries found by the \textit{Kepler} spacecraft, which lie relatively close to (and below) the Roche-lobe filling $P(m_{\rm c})$ line of AGB stars \citep{KruseAgol2014,Kawahara2018,Belloni2024_koi,Yamaguchi2024_KIC}. The high eccentricities of barium stars present a related long-standing puzzle \citep{Jorissen1998,Pols2003}. Specifically, natal white dwarf kicks due to asymmetric mass loss at the end of the AGB phase have been proposed as a possible solution \citep{Heyl2007,Izzard2010,Stone2015,ElBadryRix2018}. Whether or not similar kicks affect the eccentricities of binary millisecond pulsars remains to be tested.

\section*{Acknowledgements}

We thank N\'uria Navarro for valuable guidance in using the Hebrew University's computer cluster and for sharing her input files. We thank Re'em Sari, Nicholas Stone, and Natsuko Yamaguchi for illuminating discussions, and we are grateful to the reviewer for insightful comments which improved the paper. We acknowledge support from the Israel Ministry of Innovation, Science, and Technology (grant No. 1001572596), and from the United States -- Israel Binational Science Foundation (BSF; grant No. 2022175).  

\section*{Data Availability}

The data underlying this article will be shared on reasonable request to the corresponding author.



\bibliographystyle{mnras}
\input{shells.bbl}





\appendix

\section{Evolutionary time-scales}\label{sec:timescales}

In Section \ref{sec:orbital} we assumed that as long as giant stars fill their Roche lobe, their tidal circularization time-scale $t_{\rm circ}$ is shorter than their evolutionary time-scale, thus enforcing energy equipartition between the epicyclic motion and the convective eddies. \citet{Phinney1992} explained that only after the star detaches from its Roche lobe and $r_{\rm env}/R$ drops (when $m_{\rm env}\approx m_{\rm env}^{\rm crit}$), $t_{\rm circ}\propto(r_{\rm env}/R)^{-8}$ exceeds the evolutionary time-scale such that the eccentricity freezes. 

\begin{figure}
\includegraphics[width=\columnwidth]{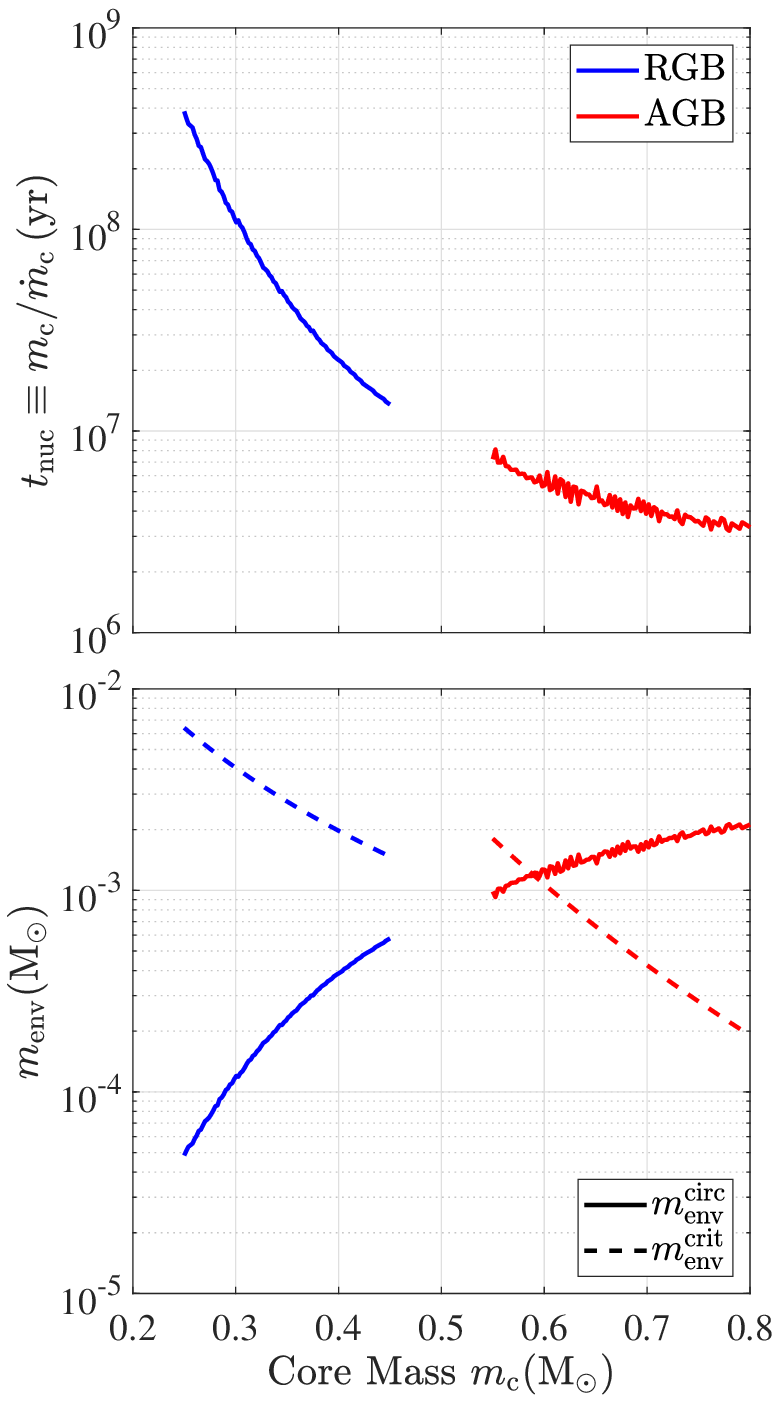}
\caption{\textit{Top panel:} the core's nuclear time-scale $t_{\rm nuc}\equiv m_{\rm c}/\dot{m}_{\rm c}$ as a function of its mass (helium core on the RGB and CO core on the AGB).
\textit{Bottom panel:} as the envelope is depleted during Roche-lobe overflow, its evolutionary time-scale $t_{\rm evo}=(m_{\rm env}/m_{\rm c})t_{\rm nuc}$ shortens whereas its tidal circularization time-scale $t_{\rm circ}\propto m_{\rm env}^{-2/3}$ lengthens, until $t_{\rm evo}=t_{\rm circ}$ when the envelope's mass is reduced to $m_{\rm env}^{\rm circ}$ (solid lines). The eccentricity freezes when $m_{\rm env}=\max(m_{\rm env}^{\rm circ},m_{\rm env}^{\rm crit})$, where $m_{\rm env}^{\rm crit}(m_{\rm c})$ (dashed lines) is given by the power-law fits in Fig. \ref{fig:mcrit}.
}
\label{fig:timescale}
\end{figure}

In Fig. \ref{fig:timescale} we check this assumption by computing the core's nuclear evolution time-scale $t_{\rm nuc}\equiv m_{\rm c}/\dot{m}_{\rm c}$, as inferred from the core's growth over time through nuclear burning in the calculations presented in Fig. \ref{fig:radius}. At the final stages of the mass transfer and after Roche-lobe detachment, the remaining envelope $m_{\rm env}$ is consumed by nuclear burning on a time-scale $t_{\rm evo}=m_{\rm env}/\dot{m}_{\rm c}=(m_{\rm env}/m_{\rm c})t_{\rm nuc}$. We compare this evolutionary time-scale to the tidal circularization time-scale assuming Roche-lobe overflow $t_{\rm circ}\propto m_{\rm c}^{1/3}m_{\rm env}^{-2/3}$ \citep{Phinney1992,GinzburgChiang2022}. As the envelope is being depleted, the evolutionary time $t_{\rm evo}\propto m_{\rm env}$ shortens, while $t_{\rm circ}\propto m_{\rm env}^{-2/3}$ lengthens. We define the circularization mass $m_{\rm env}^{\rm circ}$ as the envelope mass for which $t_{\rm evo}(m_{\rm env)}=t_{\rm circ}(m_{\rm env})$. The eccentricity freezes when $m_{\rm env}=\max(m_{\rm env}^{\rm circ},m_{\rm env}^{\rm crit})$, i.e. the eccentricity can freeze either because of contraction within the Roche lobe at $m_{\rm env}^{\rm crit}$ \citep[as assumed by][]{Phinney1992}, or because of depletion while $m_{\rm env}>m_{\rm env}^{\rm crit}$ and the giant is still filling its Roche lobe. 

By comparing $m_{\rm env}^{\rm circ}$ to $m_{\rm env}^{\rm crit}$ in the bottom panel of Fig. \ref{fig:timescale}, we see that the \citet{Phinney1992} scenario applies to the entire range of helium white dwarfs, for which $m_{\rm env}^{\rm circ}<m_{\rm env}^{\rm crit}$. However, for CO white dwarfs with $m_{\rm c}>0.6\,{\rm M}_{\sun}$, we find that $m_{\rm env}^{\rm circ}>m_{\rm env}^{\rm crit}$. Interestingly, the eccentricity of such massive white dwarfs freezes somewhat before their progenitors detach from their Roche lobe, while $m_{\rm env}>m_{\rm env}^{\rm crit}$. None the less, because of the weak dependence $e\propto m_{\rm env}^{1/6}$, this changes our estimate for the eccentricity by a factor of $(m_{\rm env}^{\rm circ}/m_{\rm env}^{\rm crit})^{1/6}\approx 1.5$ at most. In any case, $m_{\rm env}^{\rm circ}$ is much smaller than the star's original $m_{\rm env}\sim 1\,{\rm M}_{\sun}$, such that tides have sufficient time to damp any initial orbital eccentricity and establish energy equipartition.  

\begin{figure}
\includegraphics[width=\columnwidth]{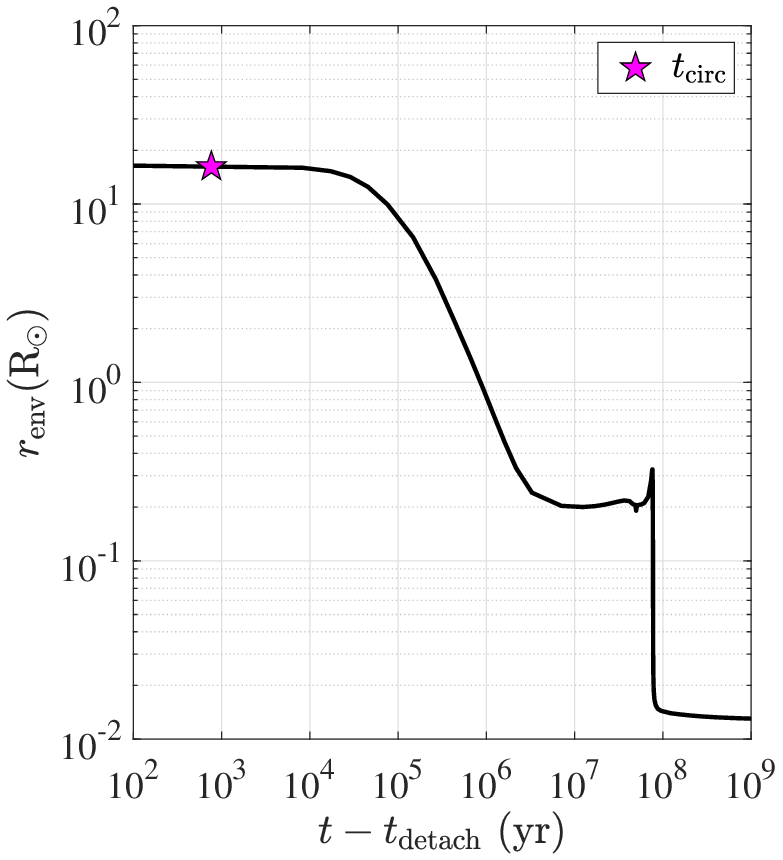}
\caption{The envelope's contraction as a function of the time elapsed since its detachment from the Roche lobe for the example presented in Figs \ref{fig:single} and \ref{fig:binary}. The circularization time-scale at the moment of detachment $t_{\rm circ}\sim 10^3\textrm{ yr}$ (magenta star marker) is orders of magnitude shorter than the contraction time-scale, similarly to helium white dwarfs \citep{Phinney1992}. 
}
\label{fig:tCaseB}
\end{figure}

In Fig. \ref{fig:tCaseB} we compare the circularization and evolutionary time-scales explicitly for the Case B overflow example of Section \ref{sec:alternative}. We find that $t_{\rm circ}$ at the moment of detachment \citep[computed as in][]{Phinney1992} is much shorter than the time-scale on which the envelope contracts, similarly to \citet{Phinney1992} and to our previous assumptions.

\bsp	
\label{lastpage}
\end{document}